# Buckling-induced quadratic nonlinearity in silicon phonon waveguide structures

Megumi Kurosu[1]*, Daiki Hatanaka[1], Hajime Okamoto[1] and Hiroshi Yamaguchi[1]

[1]*NTT Basic Research Laboratories, NTT Corporation, Atsugi-shi, Kanagawa, Japan*

E-mail: megumi.kurosu.hf@hco.ntt.co.jp

We fabricated and characterized a single-crystal silicon phonon waveguide structure with lead zirconate titanate (PZT) piezoelectric transducers. The compressive stress in a silicon-on-insulator wafer causes a membrane waveguide to buckle, leading to the quadratic nonlinearity. The PZT transducer integrated in an on-chip configuration enables us to excite high-intensity mechanical vibration, which allows the characterization of nonlinear behavior. We observed a softening nonlinear response as a function of the drive power and demonstrated the mode shift and frequency conversion. This is the first report of the nonlinear behavior caused by the quadratic nonlinearity in a buckled phonon waveguide structure. This study provides a method to control the sign and the order of nonlinearity in a phonon waveguide by utilizing the internal stress, which allows the precise manipulation of elastic waves in phononic integrated circuits.

# 1. Introduction

Phononic integrated circuits in on-chip configurations have been attracting increasing attention because elastic waves have distinct advantages such as a small velocity, short wavelength, and less radiative crosstalk compared with light waves [1-2]. Therefore, they are expected to complement photonic integrated circuits. Improving the performance of device elements in phononic integrated circuits, such as sensors, transducers, and waveguides, has been an important research target. Researchers have developed diverse platforms, including rib-type [1-3], suspended [4-5], and membrane-type [6-12] waveguides to realize phononic integrated circuits. In particular, membrane waveguides have advantages such as single-mode operation, dispersion modulation, and two-dimensional material integration.

In nanoelectromechanical systems (NEMS), mechanical nonlinearity provides useful functionalities such as mode mixing, frequency stabilization, and efficient energy harvesting in resonator structures [13-16]. Likewise, in waveguide structures, it has been theoretically proposed that useful phenomena for signal processing such as soliton waves, which propagate stably even in a nonlinear regime, and nonlinear pulse compression, which improves signal-to-noise ratio for signal detection, can be realized by controlling phonon dispersion and mechanical nonlinearity [17]. These phenomena have been demonstrated in the field of optics and used for many applications [18-19]. However, the nonlinear behavior in phonon waveguides has not been fully investigated because of the lack of transducers that can be integrated in an on-chip configuration to excite high-intensity elastic waves enough to induce mechanical nonlinearity.

In this study, we fabricated a single-crystal silicon phonon waveguide structure and employed lead zirconate titanate (PZT) as a piezoelectric transducer on a device chip. The PZT transducer enables us to excite high-intensity vibration in an on-chip configuration [20], which allows the investigation of geometric nonlinear effects in a phonon waveguide. Previously reported phonon waveguides often show a hardening nonlinear response in the power dependence of their resonance frequency due to structure elongation [8-9]. In addition, they show little quadratic nonlinearity reflecting the symmetric cross-sectional shape in the vibration direction. In contrast, the waveguide used in this study exhibits quadratic nonlinearity large enough to show the softening response. We found that the compressive stress in a silicon-on-insulator handle wafer makes the waveguides buckle after device fabrication and that the buckling configurations induce the quadratic nonlinearity. We observed the softening nonlinear response and demonstrated the mode shift and frequency conversion. This study provides a method to control the sign and order of nonlinearity in a phonon waveguide by utilizing the internal stress, which is beneficial for realizing phonon solitons and pulse compression.

## 2. Device fabrication

The silicon phonon waveguide with PZT piezoelectric transducers was fabricated using the micro-patterning procedure illustrated in Fig. 1. A PZT layer of 1 μm, $SrRuO_3$ buffered layer of 40 nm, Pt layer of 150 nm, and $ZrO_2$ layer of 60 nm, were sputtered onto a silicon-on-insulator (SOI) wafer with a high-resistivity top-silicon thickness of 400 nm, buried silicon dioxide thickness of 2 μm, and a 675-μm-thick bulk silicon substrate [Fig. 1(a)].

A dry-etching process for PZT film was used to expose the bottom Pt electrode layer [Fig. 1(b)]. A bilayer photoresist (S1813/LOR-3A) was used to increase thermal resistance, and UV-curing was performed in a nitrogen-purged environment during the post bake to increase plasma tolerance. The PZT layer was etched by inductively coupled plasma (ICP) using $BCl_3$ gas. During the ICP etching process, the Pt layer acts as a stop layer due to the high selectivity between the PZT and Pt. A two-step dry-etching process for the Pt and $ZrO_2$ layers was used to expose the Si layer [Fig. 1(c)]. The Pt layer was patterned by photo-lithography using a photomask (iP3560) and etched by ICP using $Ar/Cl_2$ (28:5) gas chemistry, and then the $ZrO_2$ layer was etched by $BCl_3$ gas. Note that, because of the low selectivity between $ZrO_2$ and Si, we were able to achieve the optimal etching time of $ZrO_2$ by measuring its etching rate. Then, a top electrode of Au/Cr (150/10 nm) was deposited on the patterned bilayer photoresist with electron-beam evaporator, followed by a lift-off process [Fig. 1(d)]. The surface area of the PZT transducer is about $660 \times 660$ μm$^2$. Elastic vibration was piezoelectrically actuated by applying AC voltage between the top and bottom electrodes.

A periodic array of ellipse air holes was fabricated in the Si layer for wet etching of the $SiO_2$ sacrificial layer [Fig. 1(e)]. Hexamethyldisilazane (HMDS) was applied prior to photoresist (S1813) spin coating to increase adhesion of the photoresist to the Si surface. The UV-cure was performed in a nitrogen-purged environment during the hardbake both to maintain the pattern shape and improve the photoresist adhesion. The Si layer was etched by reactive ion etching with $CF_4$ gas. Finally, the Si phonon waveguide structure was fabricated by etching the $SiO_2$ sacrificial layer with buffered hydrofluoric acid (BHF) for two hours [Fig. 1(f)].

Figure 2(a) shows a schematic drawing of the device, whose width 28.5 μm, and the entire length of the suspended membrane area is 6.5 mm. Top-view and cross-sectional SEM images are shown in Fig. 2(b) and 2(c), respectively, where the thickness of the membrane varies from 325 nm at the center of the membrane to 400 nm at the clamped edge. This thickness variation resulted from the finite etching rate of Si with BHF. It is important to note that the membrane is buckled due to compressive stress in the Si layer as shown in Fig. 2(c). In general, after the SOI wafer manufacturing process, the buried $SiO_2$ layer is compressively stressed due to the difference of the thermal expansion coefficient between $SiO_2$ and Si [21]. The compressed $SiO_2$ layer relaxes its structure after device fabrication, and then the thin Si membrane buckles. The buckling height of 310 nm at the center of the membrane width was measured from the cross-sectional SEM image. We calculated the compressive stress in Si layer corresponding to this buckling height by finite element method (FEM) simulation as shown in Fig. 2(d). The estimated compressive critical stress is 125 MPa, and the estimated compressive stress after fabrication is 189 MPa.

## 3. Experimental methods

In experiments, an AC voltage from a signal generator (NF, Wavefactory 1968) was amplified by a factor of 10 by a bipolar amplifier (NF, HSA 4101). The amplified AC voltage was applied to the top electrode of a piezoelectric transducer to excite elastic vibrations in a phonon waveguide structure. Because the transducer is relatively large, the whole device chip is excited at the same time. Vibrations in the waveguide were measured with an He–Ne laser Doppler interferometer (NEOARK, MLD-230V-200-NN), and the electrical output from the interferometer was filtered with a lock-in amplifier (LIA) (Zurich Instruments, UHFLI) as shown in Fig. 2(a). The bandwidth of the LIA was set at 70.56 Hz, which is equivalent to a time constant of 981.1 μs.

## 4. Results and discussion

### 4.1 Results

We measured the frequency response of the waveguide structure at a distance $L = 6$ mm from the edge on the side where the PZT transducer is located as shown in Fig. 2(a). The frequency response is shown in Fig. 3(a), where the shaded regions indicate the position of the FEM-simulated phonon bands (COMSOL Multiphysics) shown in Fig. 3(b). The pink, light blue, and yellow regions correspond to the first band (4.1-8.4 MHz), second band (7.5-17.2 MHz), and third band (14.4-20.9 MHz), respectively. The purple regions show the overlap of the first and second band and that of the second and third band. In the simulation, we assumed that the membrane is in the post-buckling configuration, where the buckling height is 310 nm, estimated from the cross-sectional SEM image. The simulated cut-off frequency is 4.1 MHz, and no phonon band exists blow this frequency. The frequency response shows low-displacement regimes near 9, 15, 20 MHz in Fig. 3(a), although phonon bands were expected to be exist. This might be because the elastic waves decay before constructive interference occurs, since these regimes are located near the band edge where the velocity of elastic waves is quite low.

As we mentioned above, the membranes buckle due to the internal compressive stress. The majority of the membranes tend to buckle in one direction; however, a few of them buckle in the opposite direction. These oppositely buckled membranes originate from fabrication variability, and they are randomly located along the waveguide length. The boundary between two oppositely buckled membranes acts as a discontinuity where elastic waves are reflected; thus, elastic domains are created as shown schematically in Fig. 4(a). The frequency response of one such domain located at $L = 1.63$ mm is also shown in Fig. 4(a). A smaller number of resonance peaks were measured compared with the frequency response shown in Fig. 3(a) because this domain includes only four periods of the unit cell.

Here, we demonstrated nonlinear mode coupling in a phonon waveguide structure. When the applied voltage was increased, a softening nonlinear response was observed, where the resonance frequency decreased as shown in Fig. 4(a). The softening response is caused by the quadratic nonlinearity, whereas a hardening response is usually observed in NEMS beams with no compressive stress or tensile stress. The hardening response is caused by the large vibration, which induces an extension of the length of the beams [22]. We consider that our softening response originates from the buckled structure as theoretically suggested [23].

For simplicity, we start with the nondimensional equation describing transverse planar nonlinear vibration of a clamped-clamped beam in the post-buckling configuration [24]:

$$\ddot{u} + u'''' + 4\pi^2 u'' - 2b^2\pi^3 \cos 2\pi x \int_0^1 u' \sin 2\pi x\, dx =$$
$$b\pi^2 \cos 2\pi x \int_0^1 u'^2\, dx + b\pi u'' \int_0^1 u' \sin 2\pi x\, dx + \frac{1}{2}u'' \int_0^1 u'^2\, dx - c\dot{u} + F(x)\cos\Omega t, \qquad (1)$$

where $u$, $t$, and $x$ are the normalized transverse displacement, the normalized time, and the normalized position along the beam, respectively. The overdot indicates the derivative with respect to the normalized time $t$, and the prime indicates the derivative with respect to the normalized position $x$. $c, F$, and $\Omega$ are the viscous damping coefficient, the amplitude of excitation, and the normalized excitation angular frequency, respectively. $b \left(= \frac{2}{\pi}\sqrt{\frac{(P-P_c)L^2}{EI}}\right)$ is the nondimensional buckling displacement, where $P, P_c, L, E$, and $I$ are the axial load, the critical load of the first buckling mode, the length of the beam, Young's modulus, and the moment of inertia of the cross section, respectively. The nonlinear response of the system is derived from Eq. (1) and then the nondimensional effective nonlinear coefficient $\alpha_{\text{eff}}$ is given by [23]

$$\alpha_{\text{eff}} = \frac{1}{8\omega}(b^2\alpha_2 + 3\alpha_3), \qquad (2)$$

where $\omega$, $\alpha_2$, and $\alpha_3$ are the nondimensional angular frequency, the quadratic nonlinear term, and the cubic nonlinear term, respectively. The nondimensional effective nonlinear coefficient consists of two terms. The first term, related to the quadratic nonlinearity, is caused by the buckling configuration and can be both positive and negative, which is derived from the first and second terms on the right-hand side of Eq. (1). The second term, related to the cubic nonlinearity, is always negative, which is derived from the third term on the right-hand side of Eq. (1) [23]. Whether a system exhibits softening or hardening behavior depends on the value of the quadratic and cubic nonlinear term. If we approximately apply the nonlinear beam theory to our waveguide by regarding it as a coupled-beam structure, it can exhibit a softening behavior due to the quadratic nonlinearity caused by the post-buckling configuration.

Nonlinear mode coupling originates from the variation of overall tension across the waveguide structure, which is caused by one mode excited with large amplitude affecting the dynamics of other modes. To investigate the effect of mode coupling, two modes were selected as the driven (DR) mode and detection (DET) mode as shown in Fig. 4(a). The detailed nonlinear frequency response of the DR mode (5.181 MHz) is shown in Fig. 4(b). The frequency response of the DET mode (5.8654 MHz) was measured when the DR mode was excited to the large amplitude. First, the DR mode was down-swept from 5.2 MHz to a target frequency with the applied voltage of 14 $V_{pp}$. Then, the frequency response of the DET mode was measured with the applied voltage of 0.7 $V_{pp}$ to determine the shift of the resonance frequency with respect to the amplitude of the DR mode displacement. The frequency shift of the DET mode is shown in Fig. 4(c) for different drive frequencies of the DR mode. The shift of resonance peaks is much larger than the spectral width of the DET mode. The resonance frequency of the DET mode decreases when the amplitude of the DR mode becomes larger as shown in Fig. 4(d). The green shaded region in Fig. 4(d) indicates that a single peak was not observed, because more complicated nonlinear phenomena might be involved due to larger amplitude of the DR mode.

Finally, nonlinear frequency conversion was performed at $L = 1.5$ mm, where three-wave-mixing induced by the quadratic nonlinearity is demonstrated. Two AC voltages were applied to the PZT transducer at the same time. One was applied at 5.181 MHz (on-resonance) or 5.168 MHz (off-resonance) ($f_1$) with the voltage of 14 $V_{pp}$, and the other was swept around the resonance peak of 6.754 MHz ($f_2$) with the voltage of 2.1 $V_{pp}$. The demodulation frequency of the LIA was set to $f_3 = f_1 + f_2$. The green lines in Fig. 4(e) show a resonance peak whose resonance frequency corresponds to 11.935 MHz (= 5.181 + 6.754 MHz). This indicates that frequency conversion is demonstrated in a phonon waveguide structure, where a new frequency, $f_3$, is generated via the quadratic nonlinearity. On the other hand, when $f_1$ was set to the off-resonance frequency, no peak was observed, because the amplitude of the off-resonance peak was not large enough to induce nonlinearity. This three-wave mixing is realized by the quadratic nonlinearity and would not be observed without the buckling configuration.

4.2 Discussion

The frequency shift of the DET mode is expected to be proportional to the square amplitude of the DR mode for weak piezo driving. The normalized frequency shift is expressed as [23]

$$\Delta\Omega_{\text{DET}} \equiv \frac{\Delta\omega_{\text{DET}}}{\omega_{\text{DET}}} = -\alpha_{\text{DET,DR}} A_{\text{DR}}^2, \qquad (3)$$

where $\Delta\Omega_{\text{DET}}$ is the fractional frequency shift of the DET mode, $\omega_{\text{DET}}$ is the linear resonance frequency of the DET mode, and $A_{\text{DR}}$ is the amplitude of the DR mode. The amplitude of the DET mode is neglected because its amplitude is sufficiently small compared with that of the DR mode. The value of mode coupling coefficient $\alpha_{\text{DET,DR}}$ is estimated to be $5.4 \times 10^{-6}$ (nm$^{-2}$) from the linear fitting using Eq. (3) as shown in the inset of Fig. 4(d). This value is the same order of magnitude as the values reported for clamped beams, but with the opposite sign [22, 25-26]. The amplitude of the DR mode in this study is one order of magnitude larger than in previous studies owing to excellent piezoelectric coefficient of the PZT transducers. This ability to induce elastic waves with large displacement is beneficial for studying novel nonlinear phenomena in a phonon waveguide. Moreover, to the best of our knowledge, this is the first observation of softening nonlinear response in a membrane phonon waveguide structure. Control of dispersion and nonlinear effects is important for manipulating elastic waves in a phonon waveguide. Although waveguide dispersion can be modified by changing the geometrical parameters of waveguides [27], no study has reported a way to change the sign of nonlinear effects of phonon waveguides. Since control of nonlinearity in a phonon waveguide is vital for manipulating of elastic waves, controlling the sign of nonlinearity with the internal stress will pave the way to expanding the functionality of phononic integrated circuits.

## 5. Conclusions

We fabricated a single-crystal silicon phonon waveguide structure with PZT piezoelectric transducers and investigated the frequency response by FEM simulation and experiment. The compressive stress in a SOI wafer causes a softening nonlinear response, and the mode shift and frequency conversion induced by nonlinear mode coupling were demonstrated.

This study proposes that the sign and order of nonlinear effects can be controlled by changing the buckling level of a phonon waveguide, which will be beneficial for making further advances in phononic circuit devices.

## Acknowledgments

The authors thank K. Nishiguchi, R. Ohsugi, and H. Murofushi for fruitful discussions.

## References

1) W. Fu, Z. Shen, Y. Xu, C.-L. Zou, R. Cheng, X. Han, and H. X. Tang, Phononic integrated circuitry and spin{orbit interaction of phonons, Nature communications **10**, 2743 (2019).

2) F. M. Mayor, W. Jiang, C. J. Sarabalis, T. P. McKenna, J. D. Witmer, and A. H. Safavi-Naeini, Gigahertz phononic integrated circuits on thin-film lithium niobate on sapphire, Phys. Rev. Applied **15**, 014039 (2021).

3) W. Wang, M. Shen, C.-L. Zou, W. Fu, Z. Shen, and H. X. Tang, High-acoustic-index-contrast phononic circuits: Numerical modeling, Journal of Applied Physics **128**, 184503 (2020).

4) Y. D. Dahmani, C. J. Sarabalis, W. Jiang, F. M. Mayor, and A. H. Safavi-Naeini, Piezoelectric transduction of a wavelength-scale mechanical waveguide, Phys. Rev. Applied **13**, 024069 (2020).

5) R. N. Patel, Z. Wang, W. Jiang, C. J. Sarabalis, J. T. Hill, and A. H. Safavi-Naeini, Single mode phononic wire, Phys. Rev. Lett. **121**, 040501 (2018).

6) D. Hatanaka, I. Mahboob, K. Onomitsu, and H. Yamaguchi, Phonon waveguides for electromechanical circuits, Nat Nanotechnology **9**, 520 (2014).

7) M. Kurosu, D. Hatanaka, K. Onomitsu, and H. Yamaguchi, On-chip temporal focusing of elastic waves in a phononic crystal waveguide, Nature Communications **9**, 1331 (2018).

8) M. Kurosu, D. Hatanaka, and H. Yamaguchi, Mechanical Kerr nonlinearity of wave propagation in an on-chip nanoelectromechanical waveguide, Phys. Rev. Applied **13**, 014056 (2020).

9) J. Cha and C. Daraio, Electrical tuning of elastic wave propagation in nanomechanical lattices at MHz frequencies, Nature Nanotechnology **13**, 1016 (2018).

10) E. Romero, R. Kalra, N. Mauranyapin, C. Baker, C. Meng, and W. Bowen, Propagation and imaging of mechanical waves in a highly stressed single-mode acoustic waveguide, Phys. Rev. Applied **11**, 064035 (2019).

11) Y. Wang, J. Lee, X.-Q. Zheng, Y. Xie, and P. X.-L. Feng, Hexagonal boron nitride phononic crystal waveguides, ACS Photonics **6**, 3225 (2019).

12) S. Kim, J. Bunyan, P. F. Ferrari, A. Kanj, A. F. Vakakis, A. M. van der Zande, and S. Tawfick, Buckling-mediated phase transitions in nano-electromechanical phononic waveguides, Nano Letters **21**, 6416 (2021).

13) R. Lifshitz and M. C. Cross, Reviews of Nonlinear Dynamics and Complexity (WileyVCH, 2008).

14) D. Antonio, D. H. Zanette, and D. L_opez, Frequency stabilization in nonlinear micromechanical oscillators, Nature Communications **3**, 806 (2012).

15) B. Andò, S. Baglio, C. Trigona, N. Dumas, L. Latorre, and P. Nouet, Nonlinear mechanism in mems devices for energy harvesting applications, Journal of Micromechanics and Microengineering **20**, 125020 (2010).

16) F. Cottone, L. Gammaitoni, H. Vocca, M. Ferrari, and V. Ferrari, Piezoelectric buckled beams for random vibration energy harvesting, Smart materials and structures **21**, 035021 (2012).

17) A. H. Nayfeh and D. T. Mook, Nonlinear Oscillations, Wiley Classics Library (Wiley, 2008).

18) G. Agrawal, Nonlinear Fiber Optics, 5th ed. (Academic Press, Boston, 2013).

19) A. Blanco-Redondo, C. Husko, D. Eades, Y. Zhang, J. Li, T. Krauss, and B. Eggleton, Observation of soliton compression in silicon photonic crystals, Nat Communications **5**, 3160 (2014).

20) P. Muralt, Ferroelectric thin films for micro-sensors and actuators: a review, Journal of Micromechanics and Microengineering **10**, 136 (2000).

21) E. Iwase, P.-C. Hui, D. Woolf, A. W. Rodriguez, S. G. Johnson, F. Capasso, and M. Lon_car, Control of buckling in large micromembranes using engineered support structures, Journal of Micromechanics and Microengineering **22**, 065028 (2012).

22) M. H. Matheny, L. G. Villanueva, R. B. Karabalin, J. E. Sader, and M. L. Roukes, Nonlinear mode-coupling in nanomechanical systems, Nano Letters **13**, 1622 (2013).

23) W. Lacarbonara, A. H. Nayfeh, and W. Kreider, Experimental validation of reduction methods for nonlinear vibrations of distributed-parameter systems: analysis of a buckled beam, Nonlinear Dynamics **17**, 95 (1998).

24) S. A. Emam and A. H. Nayfeh, On the nonlinear dynamics of a buckled beam subjected to a primary-resonance excitation, Nonlinear Dynamics **35**, 1 (2004).

25) P. Truitt, J. Hertzberg, E. Altunkaya, and K. Schwab, Linear and nonlinear coupling between transverse modes of a nanomechanical resonator, Journal of Applied Physics **114**, 114307 (2013).

26) H. Westra, M. Poot, H. Van Der Zant, and W. Venstra, Nonlinear modal interactions in clamped-clamped mechanical resonators, Physical review letters **105**, 117205 (2010).

27) D. Hatanaka, A. Dodel, I. Mahboob, K. Onomitsu, and H. Yamaguchi, Phonon propagation dynamics in band-engineered one-dimensional phononic crystal waveguides, New Journal of Physics **17**, 113032 (2015).

## Figure Captions

**Fig. 1.** Fabrication silicon phonon waveguide structure with PZT piezoelectric transducers. (a) Layer structure of the device wafer. (b) Dry etching for PZT layer. (c) Dry etching for Pt layer of the bottom electrode and $ZrO_2$ layer. (d) Top electrode and bonding pad formed by electron beam evaporation of Au and Cr. (e) Dry etching for Si layer. (f) $SiO_2$ sacrificial-layer etching with buffered hydrofluoric acid.

**Fig. 2.** (a) Schematic of the device and measurement setup. An AC voltage from a signal generator is amplified by a bipolar amplifier (BA), and the amplified AC voltage is applied to PZT transducer. Vibrations in a phonon waveguide are optically detected by laser Doppler interferometry, and the detected signals are measured using a lock-in amplifier. (b) SEM image of the phonon waveguide. The width (yellow arrow) and pitch of the ellipse hole (green arrow) are 28.5 and 10 μm, respectively. The white dashed line indicates the unit cell. The scale bar is 10 μm. (c) Cross-sectional SEM image of the phonon waveguide. The scale bar is 10 μm. (d) Simulated post-buckling configuration. The buckling height is 310 nm, estimated from the cross-sectional SEM image.

**Fig. 3.** (a) The frequency response of the phonon waveguide at 6 mm, excited by the PZT transducer with 2.8 $V_{pp}$, where the pink, light blue, and yellow regions correspond to the first band (4.1-8.4 MHz), second band (7.5-17.2 MHz), and third band (14.4-20.9 MHz), respectively. The purple regions show the overlap of first and second band and that of the second and third band. (b) FEM-simulated dispersion relation, where the internal compressive stress of 189 MPa, which corresponds to the buckling height of 310 nm, is included. Left: FEM simulated spatial profiles of the phonon modes to the different phonon branches.

**Fig. 4.** (a) Frequency responses of the phonon waveguide at 1.63 mm, where the red, blue, and green lines indicate up-sweep with 14 $V_{pp}$, down-sweep with 14 $V_{pp}$, and up sweep with 0.7 $V_{pp}$, respectively. The red arrows represent the frequency of the DR mode and DET mode. The left inset shows a schematic of domain formation at 1.63 mm, where the four periods of the unit cell are sandwiched by the oppositely buckled membranes. (b) Enlarged frequency response of the DET mode for down-sweep with 14 $V_{pp}$. (c) Frequency response of the DET mode for different frequencies of the DR mode. The applied voltages are 14 $V_{pp}$ for the DR mode and 0.7 $V_{pp}$ for the DET mode. (d) Resonance frequency of the DET mode as a function of the amplitude of the DR mode, where no single peak is observed in the green shaded region. The inset shows the fractional frequency shift of the DET mode, plotted against the squared amplitude of the DR mode. The red line shows fitting results using Eq. (3), with $\alpha_{DET,DR}$ of $5.4 \times 10^{-6}$ (nm$^{-2}$). (e) Frequency conversion due to the quadratic nonlinearity at 1.5 mm. The dashed line represents the resonance peak of $f_2$ shifted by 5.181 MHz. The green and grey lines indicate the results of frequency conversion with on- and off- resonance of $f_1$, respectively. The inset shows the frequency response around $f_1$ with the applied voltage of 0.7 $V_{pp}$, where the red and grey arrows indicate the on -and off- resonance frequency.

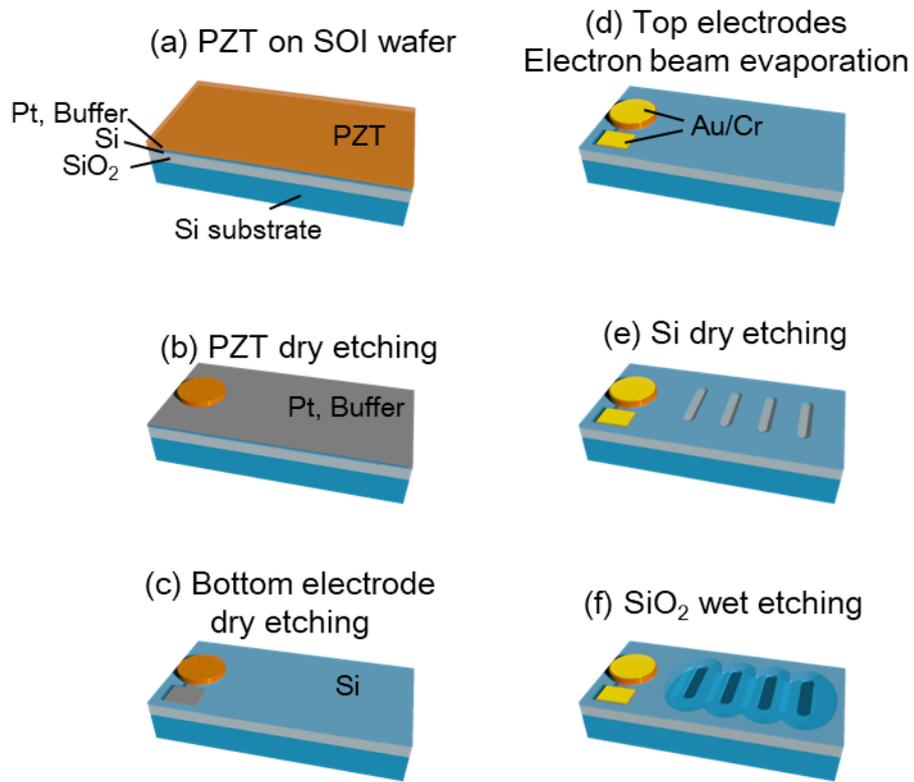

Fig. 1

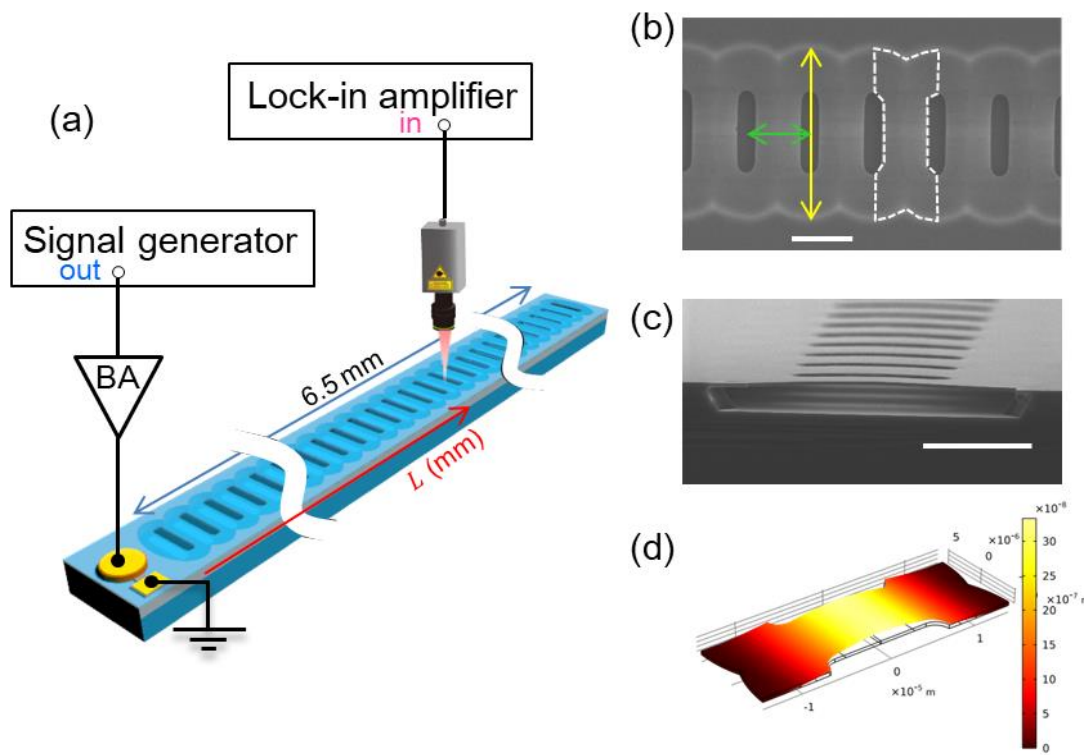

Fig. 2

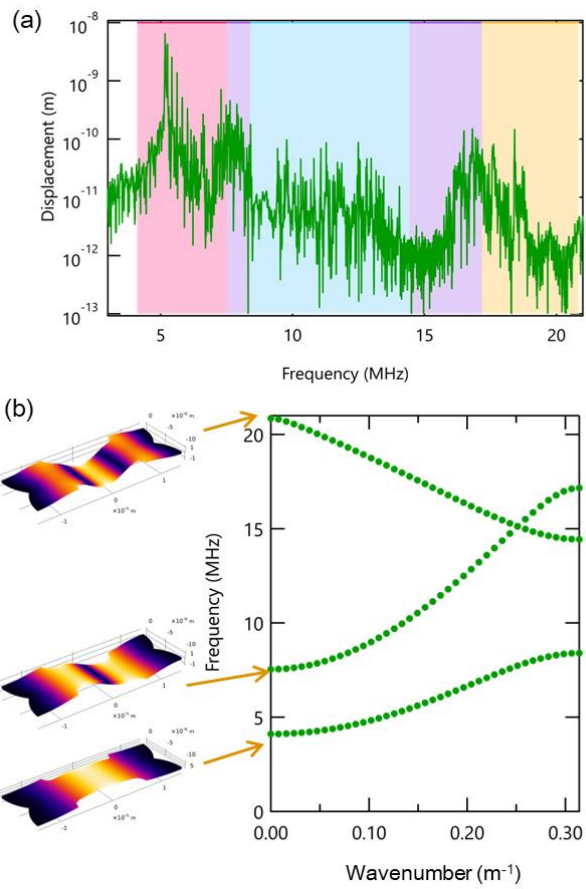

Fig. 3

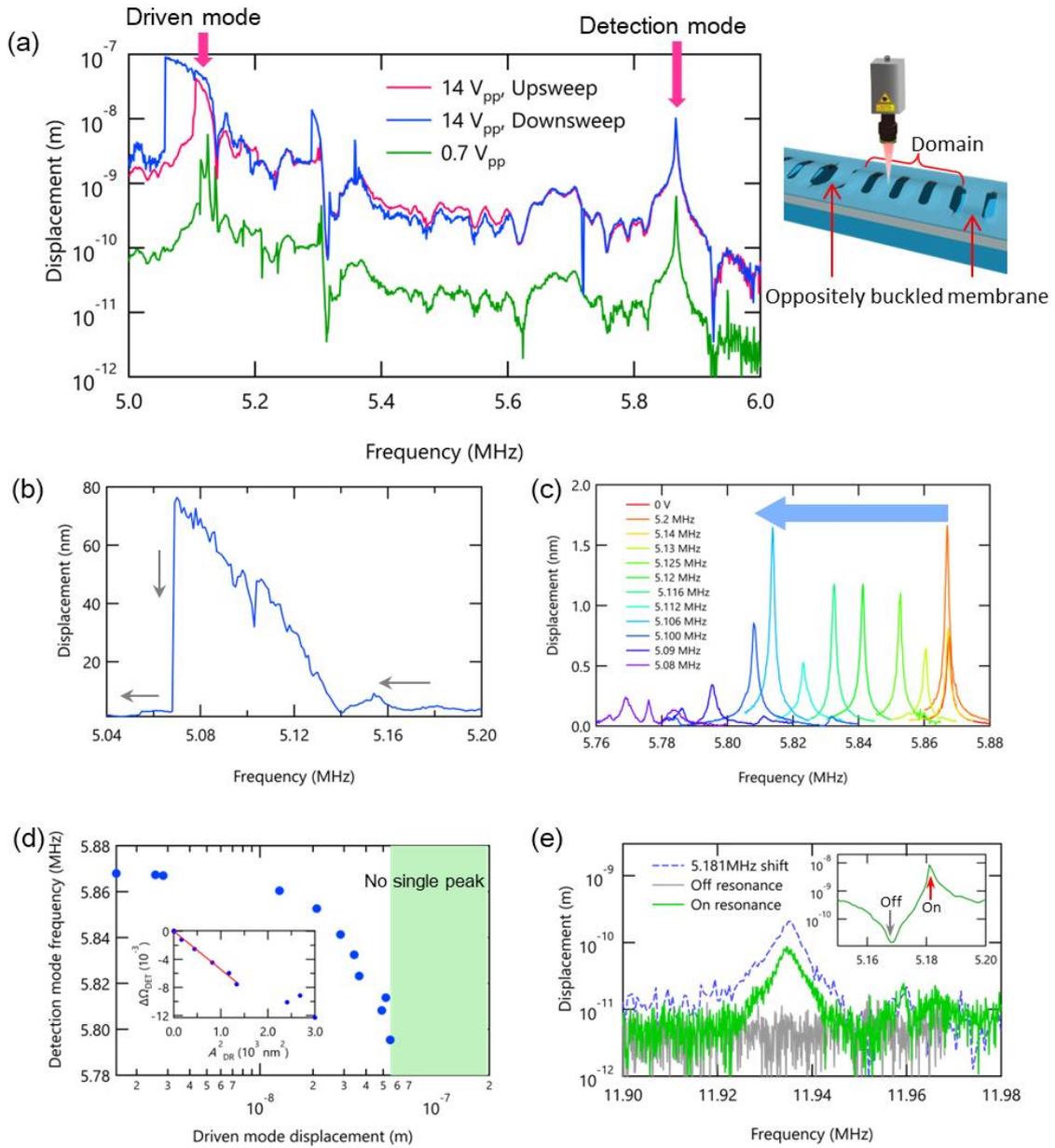

Fig. 4